\documentclass[11pt]{extarticle}

\pdfoutput=1

\usepackage{graphicx}

\usepackage{amsmath}
\usepackage{ragged2e}
\usepackage{amssymb} 
\usepackage{stmaryrd}
\usepackage{gensymb}

\usepackage{breakcites}

\usepackage{authblk}

\usepackage{geometry}
\usepackage{fancyhdr}
\makeatletter
\def\@biblabel#1{}
\makeatother

\newcommand{\citep}{\cite}

\title{Polyadic Entropy, Synergy and Redundancy among \\ Statistically Independent Processes in Nonlinear \\ Statistical Physics with Microphysical Codependence}

\author[1]{Rui A. P. Perdig\~ao\footnote{Address: TU Wien, Karlsplatz 13/222, A-1040 Vienna, Austria. E-Mail: rui.perdigao@tuwien.ac.at.}}
\affil[1]{Technische Universit\"at Wien (TU Wien), Vienna, Austria }

\date{\small \today}

\begin{document}

\maketitle

\abstract{The information shared among observables representing processes of interest is traditionally evaluated in terms of macroscale measures characterizing aggregate properties of the underlying processes and their interactions. Traditional information measures are grounded on the assumption that the observable represents a memoryless process without any interaction among microstates. Generalized entropy measures have been formulated in non-extensive statistical mechanics aiming to take microphysical codependence into account in entropy quantification. By taking them into consideration when formulating information measures, the question is raised on whether and if so how much information permeates across scales to impact on the macroscale information measures. The present study investigates and quantifies the emergence of macroscale information from microscale codependence among microphysics. In order to isolate the information emergence coming solely from the nonlinearly interacting microphysics, redundancy and synergy are evaluated among macroscale variables that are statistically independent from each other but not necessarily so within their own microphysics. Synergistic and redundant information are found when microphysical interactions take place, even if the statistical distributions are factorable. These findings stress the added value of nonlinear statistical physics to information theory in coevolutionary systems.}

\section{Introduction}

\subsection{General Motivation}

Traditional information-theoretical measures quantify the amount of macroscale information i.e. statistical properties required to characterize a system (Information Entropy, \cite{Shannon1948, CoverandThomas1991}), the amount of information shared i.e. redundant among processes (Redundant Information, e.g. the measures in \cite{CoverandThomas1991, Schneidmanetal2003}), and the amount of innovative information emerging from the non-redundant cooperation among processes (Synergistic Interaction Information, e.g. the measures in \cite{McGill1954, PiresPerdigao2015}).

The information-theoretical measures of entropy, redundancy and synergy are traditionally formulated assuming independence among the microstates living within a macroscale variable of interest. The latter is then assumed to represent a random process in which each observational event is an independent realization within a distribution embodying the statistical envelope aggregating the aforementioned microstates.
That is, while evaluating statistical relationships among macroscale variables, the microphysical interactions within at a statistical mechanical level remain elusive. If no such interactions take place e.g. in a perfect gas, the traditional information metrics will be accurate. However, that may no longer be the case when the assumption of independent microphysics no longer holds, such as in complex systems with nonlocal interactions leading to long-range codependence and multifractal scaling characteristics.

\subsection{Statistical and Information Entropy}\label{s1,2}

One core assumption in statistical physics resides in the probabilistic interpretation of Entropy. In short, that view interprets the energy densities of a system in phase space as being probabilities that the system exhibits the configurations enclosed by the volume elements where those densities are characterized. 

For systems with numerable independent microstates of cardinality $N$, this brings the original Boltzmann-Gibbs formulations to the popular definition of statistical entropy \cite{Lage1995, Callen2001}:
\begin{equation}\label{se}
S = - k \sum_{i=1}^N p_i \log (p_i)
\end{equation}
where $p_i$ is the probability associated to the microstate indexed as $i$ and $k$ is a positive constant, which in the original thermodynamic context is given by the Boltzmann constant.

In the context of Shannon's mathematical theory of communication \cite{Shannon1948}, the microstates indexed as $i$ are represented as messages characterizing a system and $p_i$ are the probabilities assigned to them. The higher the probability of a state is, the more information will be gained about the system through the associated message. 
Information Entropy is then formulated in that sense as the canonic information requirements to fully characterize the state of a system, as a function of how likely its states are. The functional form is equivalent to that of \eqref{se} with $k=1$:

\begin{equation}\label{ie}
H = - \sum_{i=1}^N p_i \log_a (p_i)
\end{equation}
For the remainder of the paper, statistical entropy shall be represented by the information-theoretical notation $H$ in order to remain clear that we are working with a scaled Boltzmann constant $k=1$ in Equation \eqref{se}.

When $a=2$, Equation \eqref{ie} corresponds to the Shannon Entropy used in computing and most information-theoretical applications, with units in $bit$. For instance, the Information Entropy of a fair coin toss is 1 bit, corresponding to the single binary (e.g. yes/no) question required to find its actual outcome. If, instead, the natural-based logarithm $a=e$ is taken, the Information Entropy is given in $nat$.

In statistical terms, the aforementioned entropies are functionally equivalent and can be denoted as BGS Entropy \cite{Kaniadakisetal2005}.

\subsection{Information Redundancy}\label{s1,3}

Information-theoretical measures of redundancy quantify the amount of information shared (and thus redundant) among a set of macroscale variables assuming that the underlying microphysics are disentangled. A redundancy-based complex system network will thus entail codependences among macroscale processes whilst the eventual codependences within microphysics are still elusive. 
 An example of such a measure is Multi-Information \cite{Schneidmanetal2003}, a non-negative quantity that can be expressed in the following form \cite{PiresPerdigao2015}:
\begin{align}
I_i(\mathbf Y) & \equiv H_i(\mathbf Y) - H(\mathbf Y) \\
& \equiv I(Y_1, Y_2, \cdots, Y_N) \\
& \equiv E \left\{ \log \left[ \frac{\rho_{\mathbf Y}}{\prod_i^N \rho_{Y_i}} \right] \right\} \geq 0
\end{align}
This measures the global statistical dependence among the components $Y_i$ of $\mathbf Y$. A well-known particular case is obtained for $N=2$, leading to Mutual Information \cite{CoverandThomas1991}:
\begin{equation}
I(Y_i,Y_j) = H(Y_i) + H(Y_j) - H(Y_i,Y_j)
\end{equation}
which interestingly has a non-null, positive lower bound (general theoretical proof in \cite{PiresPerdigao2012}, and for finite samples in \cite{PiresPerdigao2013}).

\subsection{Information Synergy}\label{s1,4}

Information-theoretical measures of synergy quantify the amount of information present in the system as a whole that is not present in any of its strict subsets. As such, synergy refers to the information gain stemming from the collective cooperation among system constituents.

One such measure is Interaction Information, which can be expressed as \cite{JakulinandBratko2004}:
\begin{align}\label{it}
I_t(\mathbf Y) = -(-1)^N \sum_{m=1}^N (-1)^m \sum_{\mathbf Z \subseteq Y} H(\mathbf Z)
\end{align}
where $m$ is the index referring to the component $Y_m$ of $\mathbf Y$.

The Interaction Information is \textit{Synergistic} across the components of $\mathbf Y$ when $I_t > 0$ for $\dim(\mathbf Y) \geq 3$. 
Conversely, when $I_t <0$ the Information is \textit{Redundant} (a negative synergy entails redundancy), becoming equivalent to multi-information.

A necessary condition for a non-redundant synergistic polyad to emerge within $\mathbf Y$ lies in the non-Gaussianity of its joint distribution. That is, from a macroscopic standpoint, only statistical moments of order $n > 2$ can enclose synergistic information. Therefore, the Interaction Information can also be expressed directly in terms of such higher-order moments. Detailed formulation details are found in \cite{PiresPerdigao2015}. 

\section{Polyadic 'Non-Extensive' Entropies}

\subsection{The Fundamentals}

Our ultimate aim is to express macroscale information theoretical measures in terms of non-extensive statistical mechanics. For that purpose, consider the following functional class of generalized entropy measures equivalent to those in \cite{BorgesandRoditi1998} and \cite{Kaniadakisetal2005}:

\begin{equation}\label{BR98H}
H_{q,r}(\mathbf Y) = k \sum_{i=1}^N \frac{p_i^r - p_i^q}{q-r}
\end{equation}
where $p_i \equiv p_{\mathbf Y}(y_i)$ is the probability of occurrence of the microstate $y_i$ in the macroscale variable $\mathbf Y$, $N$ is the number of microstates, and $k$ is a positive constant reflecting the fact that entropy can only be defined up to a multiplicative constant (i.e. only entropy differences can be fully determined).

The parameters $q$ and $r$ are associated to the existence of nonlinear codependencies among microstates within $\mathbf Y$, 
e.g. with respect to the probability of the state at which the system lies and to that of the neighbouring states \cite{CuradoandNobre2003}. In fact, the presence of these two parameters in the entropy functional stems from the dependence, relative to two different powers of the probability distribution, of the nonlinear interaction terms in the associated nonlinear Fokker-Planck Equation \cite{Schwaemmleetal2007}. Anomalous diffusion with coupling among two scaling regimes and nonlinear cross-scale coevolution in complex dynamical systems are among the physically relevant examples where this entropy functional can play an important role in eliciting macroscale thermodynamic effects of nonlinear statistical mechanics. In a coevolutionary system setting, the aforementioned parameters can be interpreted as coevolution indices \cite{PerdigaoBloeschl2014} or geometric parameters of the coevolution manifold \cite{Perdigaoetal2016}, entailing evolutionary dynamic codependence among mutually influencing observables.

Particular cases of Equation \eqref{BR98H} include the well-known Tsallis entropy \cite{Tsallis1994} by taking either $q=1$ or $r=1$ (but not both), and the seminal Botzmann-Gibbs-Shannon (BGS) entropies by taking both parameters as unity, i.e. $q=r=1$, a situation representing the absence of microphysical codependencies. 

In a joint multivariate $\mathbf Y$, say $\mathbf Y=(Y_1,\cdots,Y_M)$ of $M$ components, there can be microphysical interactions across its macroscale components or dimensions rather than solely within each marginal, which means that the joint entropy of $\mathbf Y$ will not necessarily be the sum of the entropies of its components, i.e. entropy will not be additive when there are codependencies among subsystems. Additivity only holds when microphysical interactions are confined to within each marginal or macroscale component of the multivariate $\mathbf Y$.

In order to isolate the contribution that microphysical interactions bring to the macrophysical entropy, we formulate the joint entropy of a system represented by $\mathbf Y$ built from $M$ statistically independent components or subsystems $Y_1, \cdots, Y_M$.

The a priori statistical independence of the subsystems corresponds formally to the probability factorization
\begin{equation}\label{factorp}
p_{\mathbf Y} = \prod_i^M p_{Y_i}
\end{equation}
which dictates not only bilateral but also multilateral statistical independence.

However, it does not imply factorization of the entropies, since the mixing of subsystems can introduce cross-dependencies among microstates belonging to different subsystems. Whether and under which circumstances that is the case is investigated by evaluating the difference between the entropy of the overall system $\mathbf Y$ and the sum of the entropies of its subsystems taken isolately.

\subsection{Dyadic systems}

The entropy functional in \eqref{BR98H} for a bivariate system $\mathbf Y=(Y_i,Y_j)$ where the factorization \eqref{factorp} holds is hereby derived from \eqref{BR98H} under the factorization constraint in \eqref{factorp}, yielding:
\begin{align}\label{dyadicentropy}
H_{q,r}(Y_i,Y_j) & = H_{q,r}(Y_i) + H_{q,r}(Y_j) + \\
& = (1-q)
H_{q,r}(Y_i) H_{q,1}(Y_j) + \\
& = (1-r) H_{q,r}(Y_j)  H_{r,1}(Y_i)
\end{align}
a result consistent with \cite{BorgesandRoditi1998}.

This means that the total entropy of the system [\textit{lhs} of \eqref{dyadicentropy}] is not fully assessed by the entropy of its parts when they were separate entities [first two \textit{rhs} terms of \eqref{dyadicentropy}], rather depending on nonlinear terms function of the microphysical codependence parameters $q$ and $r$.
Once coming together in the form of $\mathbf Y$, and notwithstanding their statistical independence, $Y_1$ and $Y_2$ bring about a nonlinear contribution to the joint entropy arising from the microphysical interactions represented by the parameters $q$ and $r$. When these are unit valued, the joint entropy reduces to the sum of the entropies that the parts held as separate entities.

Whether the joint entropy will be higher or lower than the sum of that from the separate components will depend on what kind of microphysical interaction is at play. For instance, when both $q$ and $r$ are lower than one, the joint entropy will be higher than the sum of the entropies of the separate subsystems (the whole will be more than the sum of the parts, entailing a dyadic synergy). When the aforementioned parameters are higher than one, the converse happens, with the entropy of the combined system being lower than the sum of the entropies of the subsystems prior to mixing, implying the existence of entropy-reducing factors, i.e. redundancy, among those subsystems.
These are then macroscale footprints of microphysical links being established across subsystems once they become connected within a bivariate whole. 

The notion that the entropy of a combined system can be lower than the sum of the entropies of its constituents taken separately can be physically understood by taking into consideration that upon combining previously separated subsystems, these can develop microphysical links resulting in a loss of dynamic freedom and hence of entropy relative to the pre-mixing stage. The connection is then captured through the nonlinear terms in the joint entropy functional. Only when no such connections exist has the entropy to be non-decreasing, which is also the case.

\subsection{Triadic systems}

The joint entropy of a triadic system $\mathbf Y=(Y_1,Y_2,Y_3)$ is hereby formulated with the entropy functional \eqref{BR98H} as follows:

\begin{align}\label{triadicentropy}
H_{q,r}(Y_i,Y_j,Y_l) & =  \sum_{\alpha \in \{i,j,k\}} H_{q,r} (Y_\alpha) + \\
& + (1-q) \left\{ H_{q,r}(Y_i) + H_{r,1}(Y_j) + \left[H_{q,r}(Y_i) + H_{q,r}(Y_j) \right] H_{q,r}(Y_l) \right\} + \\
& + (1-r) \left\{ H_{q,r}(Y_j) + H_{r,1}(Y_i) + \left[H_{r,1}(Y_i) + H_{r,1}(Y_j) \right] H_{q,r}(Y_l) \right\} + \\
& + (1-q)^2 H_{q,r}(Y_i) H_{q,r}(Y_j) H_{q,r}(Y_l) + \\
& + (1-r)^2 H_{1,1}(Y_i) H_{r,1}(Y_j) H_{q,r}(Y_l) + \\
& + (1-q)(1-r) \left[
H_{r,1}(Y_i) H_{q,r}(Y_j) H_{q,r}(Y_l) + H_{r,1}(Y_i) H_{r,1}(Y_j) H_{q,r}(Y_l)
\right]
\end{align}
where $i,j,k$ are permutations of the indices $1,2,3$ including themselves.

All nonlinear terms (16) to (20) are trivially null for BGS entropies, resulting in the additive decomposability of the joint entropy in that case, reflecting the non-existence of microphysical interactions across different subsystems.

For $q=r$, the contributions from the triadic product of entropies will always be non-negative since $(1-q)^2 \geq 0, \forall q \in \mathbb R$. This means that triadic products will always be synergistic in this regard. This is consistent with mechanistic results that triadic interactions yield statistical and dynamical synergies in complex dynamical systems, e.g. \cite{PiresPerdigao2015, Perdigaoetal2016, Perdigao2017}.

When the parameters differ ($q \neq r$), the triadic products of entropies involved in (14) admit negative outcomes, namely when $q > 1 \wedge r < 1$ or $q < 1 \wedge r > 1$. In this case, the contribution to the total entropy is negative, entailing redundancy (negative synergy).
The term (14) will always be null when $q=1$ or $r=1$, i.e. in the particular conditions associated to the Tsallis entropy. This entropy will thus entail only non-negative triadic contributions to the joint entropy i.e. capture only synergistic triads (whilst leaving out redundant, codependent ones). In doing so, the measure provides only non-redundant information, avoiding the  overestimation of information content.

\section{Polyadic Synergy and Redundancy}

\subsection{Synergy and Redundancy emerging among statistically independent variables}

With polyadic entropy functionals at hand, we are in position to quantify the synergy and redundancy associated to the nonlinear microphysical footprint onto the macrophysics (i.e. to non-trivial scaling parameters $q$ and $r$), with special interest regarding the cross-variable microphysical interactions that develop once the once-separate variables come together forming the joint system polyad.

In the current study, we consider triads involving variables $Y_i,Y_j,Y_l$ that are statistically independent a priori, i.e. for which:

\begin{equation}\label{bla}
p_{(Y_i,Y_j,Y_l)} (Y_i,Y_j,Y_l) = \prod_\alpha p_{Y_\alpha} (Y_\alpha) \, \alpha \in {i,j,k}
\end{equation}

Under these conditions the Mutual Information (MI), a known measure of statistical redundancy, is null by definition. MI characterizes statistical redundancy at the macroscopic level, i.e. information shared by probability distributions. As such, while it has a thermodynamic (macroscale) relevance, it does not inform on whether the microphysics are independent as well.

The Polyadic Synergy $\mathcal S_M$ among statistically independent subsystems $Y_\alpha, \alpha=\{i_1, \cdots, i_M\}$ enrolling in a combined $M$-component system $\mathbf Y$ is hereby defined as the difference between the sum of entropies of each subsystem prior to combination (total entropy of a juxtaposition of separate subsystems) and the entropy of the combined system wherein the subsystems are allowed to communicate (i.e. to interact) at the microphysical level, whilst retaining their macroscale statistical independence by preserving factorability as in \eqref{bla}.

Mathematically, this definition is hereby expressed as:

\begin{equation}\label{synergy}
\mathcal S_M(Y_\alpha) \equiv H_{q,r}(\mathbf Y) -  \sum_\alpha H_{q,r} (Y_\alpha)
\end{equation}

This definition leads to a trivially null synergy for the BGS entropies, which is natural since they do not admit microphysical interactions within the macroscale distribution, i.e. they assume the variables under evaluation to be memoryless random processes.

In general, the synergy will be positive when the entropy of the combined system $\mathbf Y$ exceeds the sum of the entropies of the intervening components $Y_\alpha$. Physically, this means that, upon combination, the $Y_\alpha$ cooperate to produce emerging dynamics not present in any of them a priori.

An example of a process with emerging dynamics is triadic wave resonance, wherein from a pair of statistically independent primary waves $\omega_i$ and $\omega_j$ of frequencies $f_i$ and $f_j$ respectively, a secondary wave $\omega_l$ emerges with frequency $f_l = f_i+f_j$, whilst preserving statistical independence between $\omega_i$ and $\omega_j$. The spectrum of the resulting wave system is richer than the sum of the spectra of the primary waves, thus requiring more information to characterize it. The spectral entropy of the resulting system is thus necessarily higher. Note that in this case the Equation \eqref{bla} does not hold since the waves are pairwise independent but not triadically.

Conversely, the synergy will be negative when the entropy of $\mathbf Y$ is lower than the sum of the parts $Y_\alpha$. At first sight, it might appear to be thermodynamically counter-intuitive. However, such negative synergies can emerge if there are redundancies or constraints developing from the establishment of microphysical links across the once-separated parts. Such links entail freedom loss in the dynamics, with associated reduction in the total entropy relative to the a priori detached configuration. Note also that a \textit{Negative Synergy} is fundamentally \textit{Redundancy}, consistent with the information-theoretical counterparts presented in sections \ref{s1,2}-\ref{s1,4}.

By imposing that the probability distribution of the combined system remain factorable as in \eqref{bla}, macroscale information-theoretical measures of redundancy vanish as noted above, therefore any emerging redundancy can be attributed to the microphysical interactions, the macroscale footprint of which is carried in the entropy terms involving the parameters $q$ and $r$.

In order to illustrate the explicit role of these in the synergy and redundancy among independent variables, we consider the two and three-component forms (dyadic and triadic respectively).

\subsection{Dyadic form}

The dyadic synergy $\mathcal S_2$ naturally follows from Equation \eqref{synergy} as:

\begin{equation}\label{dyadicsynergy}
\mathcal S_2(Y_i,Y_j) = H_{q,r}(Y_i,Y_j) - H_{q,r} (Y_i) - H_{q,r} (Y_j)
\end{equation}

By decomposing the joint entropy as in Equation \eqref{dyadicentropy}, $\mathcal S_2$ becomes:

\begin{equation}
\mathcal S_2(Y_i,Y_j) = (1-q) H_{q,r}(Y_i) H_{q,1}(Y_j) + (1-r) H_{q,r}(Y_j)  H_{r,1}(Y_i).
\end{equation}

Particular forms of $\mathcal S_2$ of interest are trivially obtained for entropy functionals under the Tsallis entropy, where either $q=1$ or $r=1$ (but not both), leading to:

\begin{equation}
\mathcal S^{[\mathcal T]}_2(Y_i,Y_j) = (1-q) H_{q,1}(Y_i) H_{q,1}(Y_j)
\end{equation}
or, equivalently,
\begin{equation}
\mathcal S^{[\mathcal T]}_2(Y_i,Y_j) = (1-r) H_{1,r}(Y_j)  H_{r,1}(Y_i).
\end{equation}
which means that when the sole parameter $q$ or $r$ is lower [higher] than 1 there is a positive [negative] synergy between $Y_i$ and $Y_j$ when they combine to form $Y_{i,j}$. The notational superscript $[\mathcal T]$ is intended to stress that this is the particular case under the Tsallis entropy functional.

The trivially null case then occurs in the BGS entropy functional case, leading to the known result that neither synergistic nor redundant information can be found among memoryless statistically independent processes.

Overall, whether the synergy will be positive or negative will depend not only on the entropies per se but crucially so on the relative weight of the parameters representing nonlinear microphysical interactions. A process mixing that is not captured at all on traditional information-theoretic metrics and leads to a diagnostic of positive synergy under the Tsallis entropy functional (e.g. $q<1$) may actually entail redundancy if the second parameter outweighs the cooperative (synergistic) role of $q$.

\subsection{Triadic form}

Similarly to the dyadic form, the triadic synergy associated to the joint yet factorable $Y_i,Y_j,Y_l$ relative to its disjoint a priori terms is hereby obtained by taking $M=3$ in Equation \eqref{synergy}, and decomposing the triadic entropies as in Equations (13)-(18), yielding:

\begin{align}\label{triadicsynergy}
\mathcal S_3 (Y_i,Y_j,Y_l)
& = H_{q,r}(Y_i,Y_j,Y_l) -  \sum_{\alpha \in \{i,j,k\}} H_{q,r} (Y_\alpha) = \\
& = (1-q) \left\{ H_{q,r}(Y_i) + H_{r,1}(Y_j) + \left[H_{q,r}(Y_i) + H_{q,r}(Y_j) \right] H_{q,r}(Y_l) \right\} + \\
& + (1-r) \left\{ H_{q,r}(Y_j) + H_{r,1}(Y_i) + \left[H_{r,1}(Y_i) + H_{r,1}(Y_j) \right] H_{q,r}(Y_l) \right\} + \\
& + (1-q)^2 H_{q,r}(Y_i) H_{q,r}(Y_j) H_{q,r}(Y_l) + \\
& + (1-r)^2 H_{1,1}(Y_i) H_{r,1}(Y_j) H_{q,r}(Y_l) + \\
& + (1-q)(1-r) \left[
H_{r,1}(Y_i) H_{q,r}(Y_j) H_{q,r}(Y_l) + H_{r,1}(Y_i) H_{r,1}(Y_j) H_{q,r}(Y_l)
\right]
\end{align}
where $i,j,k$ are permutations of the indices $1,2,3$ including themselves.

Again here, the synergy is null under the BGS entropies, as all of its terms vanish when both parameters are unity $q=r=1$. 

The triadic synergy for the one-parameter Tsallis entropy case then becomes:
\begin{align}\label{triadicsynergytsallis}
\mathcal S^{[\mathcal T]}_3 (Y_i,Y_j,Y_l)
& = H_{q,1}(Y_i,Y_j,Y_l) -  \sum_{\alpha \in \{i,j,k\}} H_{q,1} (Y_\alpha) = \\
& = (1-q) \left\{ H_{q,1}(Y_i) + H_{1,1}(Y_j) + \left[H_{q,1}(Y_i) + H_{q,1}(Y_j) \right] H_{q,1}(Y_l) \right\} + \\
& + (1-q)^2 H_{q,1}(Y_i) H_{q,1}(Y_j) H_{q,1}(Y_l)
\end{align}
by taking $r=1$ in Equation \eqref{triadicsynergy}. An equivalent expression is obtained by taking $q=1$ instead.
For $q \neq 1$ there will always be a positive contribution from the triadic product to the synergy in Equation \eqref{triadicsynergytsallis}. Whether or not the overall synergy will be positive or negative (redunancy) will be dependent on the behaviour of the lower-order terms.

\section{Concluding Remarks}

The formulation of generalized entropy functionals where microphysical nonlinearities are taken into account (albeit for now in a parametric manner) is relevant for the evaluation of entropy functionals in systems the microphysics of which are nonlinearly codependent, e.g. for anomalous diffusion, nonlinear coevolution and mixing among heterogeneous yet interconnected media e.g. across a boundary permeable to momentum and/or heat transfer.  
While a diversity of studies have explored such functionals and their associated mathematical properties and physical consistency, the evaluation of higher-order functionals had been largely elusive, along with the explicit formulation of the synergies arising from bringing separate subsystems together into an interconnected whole. A new measure of polyadic synergy has been introduced and discussed that takes into account the microphysical codependence through the entropic parameters present in generalized "non-extensive" entropies. These effects have been isolated by considering the synergy among statistically independent variables. As expected, when the microphysics are uncoupled the synergy is null in line with the classical information-theoretical results valid for i.i.d.~variables. However, when codependence exists within the microphysics, and albeit the macroscale statistical independence (via the associated probability factorability), a synergistic term emerges in the macrophysics, expressing the macroscale footprint of interconnected microphysics - cooperant/constructive when the synergy is positive, constraining/redundant when the synergy is negative.

Fundamentally, we can interpret the non-additivity of generalized entropies as coming down to the comparison between pre and post-mixing subsystems. If we sum the marginal entropies of all components already involved in a composite system, then they will naturally add up to that of the overall system, because entropy is fundamentally extensive at each stage of the system evolution. We can further interpret the non-extensiveness as arising from comparing system states at different stages of their evolution. In fact, the entropy of each component that will enter a system can change upon involvement in that system, even if the macroscale statistics were initially independent, since microphysical interactions across the sub-system divide can now take place that were not the case when the subsystems were not communicating (i.e. when they were separate).

The present study formulated generalized information theoretical metrics beyond the traditional assumption of memoryless microphysics, by taking nonlinear statistical physics into account. As such, it is aimed at sharing basilar ideas, concepts and developments, igniting discussion and opening windows of opportunity for further exploration by the community. 
Follow-up studies shall delve on further properties and applications to the characterization of information metrics and predictability among processes exhibiting non-trivial internal memory within their microphysics.

All in all, the study brings out the following take-home messages:
\begin{itemize}
\item
Factorable probabilities do not necessarily lead to additive entropies.
\item Microscale codependence does not necessarily lead to macroscale codependence.
\item Macroscale independendence does not necessarily imply microscale independence.
\end{itemize}
This is consistent with the knowledge that statistical independence does not imply dynamic independence \cite{Perdigaoetal2016}. Moreover, the findings of the present study stress the relevance of taking nonlinear microphysical interactions into account when formulating information-theoretical measures, especially when a system is undergoing mixing among subsystems such as in thermodynamic coevolutionary settings.

\vspace{12pt}
\paragraph{Acknowledgements:}

\textit{
The financial support from the ERC Advanced Grant 'Flood Change' project no. 291152, and from the Meteoceanics research project MR-220617 'Mathematical Physics of Complex Coevolutionary Systems', is gratefully acknowledged.
}

\end{document}